\input epsf.tex
\documentstyle[12pt]{article}
\newcommand{\be}{\begin{equation}}   \newcommand{\ee}{\end{equation}}
\newcommand{\bear}{\begin{eqnarray}}
\newcommand{\eear}{\end{eqnarray}}
\newcommand{\ba}{\begin{array}}      \newcommand{\ea}{\end{array}}

\newcommand{\ie}{{\it i.e.\ }}

\begin{document}
\pagestyle{empty}
\begin{titlepage}

\vspace*{-8mm}
\noindent 
\makebox[11.5cm][l]{UCTP 109/98} hep-ph/9806446 \\
\makebox[11.5cm][l]{} June 22, 1998 \\


\vspace{2.cm}
\begin{center}
  {\LARGE {\bf  Chiral Symmetry Breaking in $2 +1$ dimensions due to Sphalerons
 }}\\

\vspace{42pt}

Indranil Dasgupta \footnote{e-mail  address: dgupta@physics.uc.edu}\\
and\\
L. C. R. Wijewardhana \footnote{e-mail address: rohana@physics.uc.edu}\\
\vspace*{0.5cm}

 \ \ Department of Physics, University of Cincinnati \\
{400 Geology/Physics Building, Cincinnati, OH 45221, USA}

\vskip 3.4cm
\end{center}
\baselineskip=18pt

\begin{abstract}
{\normalsize {
In $2+1$ dimensional gauge thories with $SU(N_c)$ color and $2N_f$ flavors of quarks
in the fundamental representation, sphalerons, if present
may lead to quark condensation
and chiral symmetry breaking. The effect is similar to instanton induced 
chiral symmetry breaking in $3 + 1$ dimensions and is due to the interaction
of quark propagators with sphalerons. The existence of sphalerons requires
that color symmetries be broken by a Higgs in the fundamental representation. 
We show that the sphaleron effect may persist for arbitrarily large $N_f$
but vanishes along with the mass of the gauge fields corresponding to broken
generators of $SU(N_c)$. The effect is inherently non-Abelian and absent for 
QED.
 }}

\end {abstract}

\vfill
\end{titlepage}

\baselineskip=18pt  
\pagestyle{plain}
\setcounter{page}{1}



\section {Introduction}

Understanding the nature of chiral symmetry breaking in quantum chromo-dynamics (QCD) is one of the
more difficult problems in non-Abelian gauge theories. 
There are analogous but simpler problems in $2+1$ dimensions \cite {anw} where
usually one writes a gap equation 
for the quark propagator in the presence of gluon exchange or four-fermi interactions. 
A shift in the pole of the quark propagator to a non-zero value signifies chiral symmetry 
breaking. In $3+1$ dimensions a completely different non-perturbative approach
may also be taken, namely, semiclassical expansion about instantons, which may give effects comparable
to gluon exchange \cite {cdg, as}. Instantons, technically, are not present in $2+1$ dimensional gauge
theories. Nevertheless one can ask if there is a semiclassical effect paralleling the 
instanton effect of $3+1$ dimensions. 

In this paper we show that in $2+1$ dimensions, a new non-perturbative mechanism
can lead to chiral symmetry breaking.
The new mechanism is an interaction between quarks and instanton-like
field configurations
that, in the context of $3+1$ dimensional theories, would be called sphalerons. 
The relevance of
instantons to the problem originates from their being the least action gauge field configurations in 
4- dimensions that suport a normalizable quark zero mode. Sphalerons, though being different from 
instantons in many respects, share this vital property. 

The existence of the sphaleron requires that the color symmetries be broken. 
Upon integrating out the massive gauge and Higgs fields one obtains a quark action with four-fermi
(vector-vector or Thirring type) 
effective interactions. Recent studies indicate 
that if this interaction is strong enough and the number of flavors is small, quarks can condense. 
However the sphaleron effects are distinguished from gluon exchange effects in two respects. 
Firstly, we can consider a case where 
the mass of the Higgs $M_H \sim \sqrt {\lambda v^2}$ is large ($\lambda $ is a dimensionful coupling and 
the vacuum expectation value of the Higgs is $v/ \sqrt {2}$) and the gauge field is heavy ($M_W \sim gv$ where
$g$ is the gauge coupling). For small $g/v$, the low energy
four-fermi theory is weakly coupled. Four-fermi interactions can not consistently break flavor
symmetries in this theory, but sphaleron effects can. Secondly, we find that 
sphaleron effects may persist to arbitrarily large values of flavor ($N_f$).

In section 2 we explain the notation and review
the sphaleron solution. In section 3 we solve a gap equation for sphaleron- induced quark mass. In section
4 we explain the significance of the sphaleron in 3- dimensional physics and discuss some generalizations.

\section {The $SU(2)$ Model}

Let us consider the case $N_c = 2$. The Euclidean action is
\bear
S &=& \int d^3 x \left [ {\cal L}_A + {\cal L}_F + {\cal L}_S \right ] \nonumber \\
{\cal L}_A &=& {1 \over 4 } (F_{\mu \nu}^a )^2 \nonumber \\
{\cal L}_F &=& { \bar \psi_{\alpha a}} i \gamma _{\mu} D_{\mu} \psi _{\alpha a} \nonumber \\
{\cal L}_S &=& |D_{\mu}\phi |^2 + U(|\phi |)
\label {action}
\eear 
where $a =1, 2$ is the color index and $\alpha = \pm 1, .... \pm N_f$ is a flavor index. 
The scalar field $\phi $ is a doublet under
the color $SU(2)$ and $U$ is a potential, $F_{\mu
\nu}^a = \partial _{\mu} W_{\nu}^a - \partial _{\nu} W_{\mu}^a 
+ g \epsilon ^{abc} W_{\mu}^b W_{\mu}^c$
is the $SU(2)$ field strength, $W_{\mu} = W^a_{\mu}\tau _a$ are the $SU(2)$
gauge fields defined using the Pauli matrices $\tau _a$ and the covariant derivative
is $D_{\mu} = \partial _{\mu} - {1 \over 2} igW_{\mu}$. 
The Euclidean $\gamma $ matrices are $2 \times 2$
and anti-Hermitian. 
Gauge and global symmetries prevent Yukawa couplings from being generated
perturbatively.  

Let us summarize what is known about the chiral symmetries in this model. If the scalar sector
was absent then a $1/N_f$ expansion would show that the $U(2N_f)$ flavor symmetry is broken 
to $U(N_f) \times U(N_f)$ and quarks $\psi _{\pm \alpha a}$ acquire masses $\pm m$
when $N_f$ is smaller than a critical value $N_{fc}$ \cite {anw}. (The term 
chiral symmetry is appropriate precisely because of this pairwise generation 
of masses of opposite signs for the quarks). This pattern of masses preserves the action under 
a parity transformation described by:
$\psi_{\alpha} (x) \to \sigma_2\psi_{-\alpha} (x^{\prime}) \; ; 
W_{\mu}(x) \to P_{\mu}W_{\mu}(x^{\prime}) \; ; 
x^{\prime}_{\mu} = P_{\mu}x_{\mu} \; ; 
P_1 =  -P_2  = P_3 = 1$.
For the color group $G$, $N_{fc} = [{128 \over 3}  C_2(R)]/[T(R) {\pi ^2}]$, 
where $C_2(R)$ is the quadratic Casimir invariant in representation $R$
and $T(R) = C_2(R)d(R)/r$, with $d(R)$ being the dimension of the representation
$R$ and $r$ the dimension of the group $G$ \cite {anw}. 
For  $SU(N_c)$ in the fundamental represantation, $C_2(R) = (N_c^2 -1)/N_c$ and
$T(R) = 1/2$. 
The scale of the quark masses is  $g^2$, although the
masses are exponentially small for $N_f \to N_{fc} -$. 
For $N_f > N_{fc}$, there is no flavor symmetry breaking and the quarks are massless. For $SU(2)$ color
$N_{fc} = 64/\pi ^2$. 

The summary above is likely to be qualitatively unchanged upon including the scalar field if the scalar field is heavy,
and the color symmetries are not  broken.
On the other hand, if $U$ has a non-trivial minimum, color symmetries are broken and some of the gauge fields
acquire mass $\sim gv$, where $g$ is the gauge coupling and $v$ is the vacuum expectation value (VEV) 
of the Higgs. Perturbatively, 
the low energy theory then consists of the light quarks
which can condense due to the Thirring type four-fermi interactions that are obtained 
by integrating out the massive vector bosons. 
Several approaches to Thirring models of this kind have been discussed whose
results often contradict each other \cite {thirring}. 
However, the model we are interested in is solidly grounded on an underlying
gauge theory, where the Dyson-Schwinger equations for the quark two 
point function can be easily written down in a $1/N_f$ expansion. For instance,
the Dyson-Schwinger equation for the quark self energy $\Sigma (p)$ can be expanded in $1/N_f$ 
for our case exactly as in ref.~\cite {anw} with
a massive gauge field propagator. 
\be
\Sigma (p) = {2 C_2(R)\alpha \over N_f} {\rm {Tr}}\int {d^3 k \over (2\pi )^3} 
{\gamma^{\mu}D_{\mu \nu}(p-k)\Sigma (k) \gamma^{\nu} \over k^2 + \Sigma^2(k)} \, \, ,
\label {dsone}
\ee
where the leading order (in $1/N_f$) gauge field propagator is
\be
D_{\mu \nu} (p-k) = {g_{\mu \nu} - \left [{1 - \xi}\right ] 
[(p-k)_{\mu}(p-k)_{\nu}]/[k(p-k)^2 + {M_w^2 \xi \over (1 + \Pi (p-k))}] \over
(p-k)^2 [ 1 + \Pi (p-k)] + M_w^2 } \, \, ,
\label {dmunu}
\ee
and where the non-local gauge $\xi $ must be chosen to make the quark wave function 
renormalization zero \cite {nash, gcs}. 
The the lowest order approximation in $1/N_f$ gives $\Pi (p-k) = T(R) \alpha /|p-k|$ and 
$\alpha$ is defined as $N_fg^2 /8$. The gauge field mass $M_w$ is given by $M_w^2 = 8v^2\alpha /N_f$.
In the $1/N_f$ expansion, $\alpha$ and $M_w$ are to be kept fixed as $N_f$ is made large. 
In solving the gap equation one makes the approximation that the relevant physics comes
from small momenta and $\alpha /|p-k| >> 1$. In this approximation
the gauge fixing parameter $\xi $ (found by equating the quark wave function renormalization
to zero) is known to approach $2/3$  when $M_w \to 0$ \cite {nash}. Using the same approximation
one can show that $\xi \to 1$ when $M_w^2 /(\alpha |p-k|) >>1$. Thus $\xi $ is usually a 
momentum dependent function that interpolates between the constant values $2/3$ and $1$ as 
the mass $M_w$ increase from $0$ to $\alpha$. We will make the approximation that $\xi =2/3$
for small $M_w$ and $\xi = 1$ for $M_w \sim \alpha$. 
The angular integration
in (\ref {dsone}) can be exactly performed to get
\be
\Sigma (p) = {2 C_2(R)\over T(R)\pi ^2N_f p} \int dk {\Sigma (k) k
\over (k^2 + \Sigma ^2 (k))} I_0 
\label {dstwo}
\ee
where
\be
I_0 = \left [ {16 \over 3 } + {20 M_w^4 \over 27\alpha^2 (p^2
-k^2)}\right ] {\rm {Min}}(p,k) - \left [2{M_w^2 \over \alpha
}{\rm {log}} \left ({|p+k| \over |p-k|} \right ) \right ]
\label {izero}
\ee
The gap equation is difficult to solve for arbitrary $M_w$. But 
In the limit $M_w^2 / \alpha p \to 0$ one can make the further
approximation that only values of $k$ lying in the interval 
$[\Sigma (k), \alpha ]$ contribute in the gap equation. Then for
for $p << \Sigma (0)$,  $I_0$
can be replaced by the expression 
\be
I_0^{\prime} = {16 \over 3} \left ( 1 - {5 \over 36} { M_w^4 \over
\alpha^2 (k^2)}\right ) {\rm {Min}}(p,k) \, .
\label {izeroprime}
\ee
In this approximation equation (\ref {dstwo})
reduces to the gap equation in \cite
{nash} with a small perturbation proportional to the fourth power of $M_w$. The corresponding
critical number of flavors is less than $64/\pi ^2$ by a number of order
$M_w^4 / (\alpha ^2 \Sigma (0)^2)$. 
Qualitatively, this result agrees with some recent studies of the Thirring
model where the four- fermi- coupling $G$ is analogous to 
$\alpha \over M_w^2$ in our case. As $\alpha \, G$ decreases, $N_{fc}$ appears to decrease to 
zero \cite {kondo}. 

Keeping in mind that in the $1/N_f$ expansions mentioned above, the condensation takes place
due to gluon exchange forces or forces arising from four-fermi contact interactions, we now turn
to a different effect, namely the effect of sphalerons, which is non-perturbative in both 
$M_w/\alpha$ and $1/N_f$. 
The phenomenon is interesting because 
if a dynamical mass is generated by sphalerons, a residual
mass may be present even when the number of flavors exceeds the critical value $N_{fc}$ or the effective four-fermi
coupling strength $\alpha / M_w^2$ is smaller than the critical value (for a given $N_f$) obtained from 
previous studies. 

Let us briefly review the salient features of the sphaleron. To make contact with existing literature
we choose $U = \lambda (|\phi | ^2 -v^2 /2)^2$. 
A symmetric {\it ansatz} for the sphaleron is the following \cite {sphaleron}.
\bear
W_{\mu} &=& - {2i \over g} f(R) (\partial _{\mu} U^{\infty})(U^{\infty})^{-1} \nonumber \\
\phi &=& {v \over \sqrt {2}} h(R ) U^{\infty} \left (\begin {array} {c}  0 \\ 1 \end {array}
\right ) \, 
\label {sphalform}
\eear
where the $SU(2)$ elements $U^{\infty}$ are defined as $U^{\infty} = 
{i{\vec {\tau}} \cdot \hat {\bf {x}}}$ and all fields are functions of the dimensionless radial coordinate
$R \equiv vgr$ where $r^2 \equiv x_1^2 + x_2^2 + x_3^2$.
There exist non-trivial solutions for $h(R)$ and $f(R)$ 
(not known in closed form) that satisfy the equations of motion. Note that this {\it ansatz} is in a manifestly
$SO(3)$ symmetric gauge. 

Historically
the sphaleron originates from being a saddle point of the energy functional in a $3+1$ dimensional gauge
theory. In that context it is a 3- dimensional field configuration at the ``top'' (the height measures energy
in $3+1$ dimensions and action in $2+1$ dimensions)
of a non-contractible loop (NCL) 
in the configuration space (where the ``bottom'' of the loop is the vaccum configuration). Its existence
therefore is due to a topological reason resembling  that of the instantons in gauge theories. Yet, being at the
``top'' of the NCL, it is, in the present context, a saddle point of the action and 
unlike the usually encountered instantons, is not a local minimum of the action. 
Nevertheless, the sphaleron is believed to have only one direction of instability (\ie  the energy/action decreases
in only one direction). 
For our purposes the most important property of the sphaleron is that 
it has quark zero modes. 

Non-perturbative contributions to quantum amplitudes due to the sphaleron can be calculated
by a saddle point expansion around the sphaleron. Multiple sphaleron effects are most easily
taken into account by considering a dilute gas of non-interacting sphalerons. 
For instance the contribution of sphalerons
to the vacuum to vacuum transition amplitude is:
\be
<0|0> = \sum _n <0|0>_n = \sum _n {\left [ J\, {\rm {exp}} [-S_S] d^{N_f} [{{m_0} \over (vg)}]^{2N_f} [\int d^3 x 
(vg)^3] \right ]^n \over n!} \, 
\label {vacuum}
\ee
The r.h.s is the sum of contributions from $n$ non-interacting sphalerons. 
The contribution of a single sphaleron is raised to the $n$-th power and divided by the symmetry factor
$n!$. Various terms in equation (\ref {vacuum}) can be understood as follows.
$S_s$ is the 
action of the sphaleron and $d$ is the determinant coming from (properly renormalized) 
Gaussian integrals over a pair of quarks of opposite chirality ($\psi _{\alpha}$ and $\psi_{-\alpha}$)
in the sphaleron background. In computing this
determinant the zero-modes of the operator $i\gamma D$ are not included. The contribution of these
zero modes is contained in the factor $m_0^{2N_f}$, where $m_0$ is the expectation value of the quark mass
operator in the zero eigenmode $\psi_{0}$ (summed over the color index $a$)
\be
m_0 = <\psi _{0 a} | m | \psi_{0 a}> \, .
\label {mzeroone}
\ee
Note that if no dynamical mass is generated for the quarks ($m_0 = 0$), the whole amplitude collapses
to zero. The factor $J$ contains 
properly renormalized determinants coming from Gaussian integrals over bosonic and ghost fields and Jacobian 
factors obtained after converting the final integral over translational zero modes of fields to the
collective coordinate $x$ denoting the location of the sphaleron. 
\be 
J = {\rm {Det}}^{\prime}[-D^2 - U^{\prime \prime}]^{-1/2} {\rm {Det}}^{\prime}[-D^2 - 2F]^{-1/2}
 {\rm {Det}}[-D^2] M 
\label {jay}
\ee
where the sphaleron background fields are to be used in computing $D$ and $U$. Each determinant is understood 
to be renormalized by division with corresponding determinants with $D$ and $U$ computed with trivial
background fields. 
The prime on some of the determinants implies omission of zero eigenvalues from translation and gauge rotation and omission
of the negative eigenvalue associated with the bosonic determinants. The negative eigenvalue appears because the sphaleron
is not a local minimum of the action but a saddle point. Such a negative eigenvalue appears also in connection with 
saddle points (called bounces) that are associated with vacuum tunneling. There, one usually completes the Gaussian integration
over the negative eigenmode by analytic continuation, obtaining an imaginary part in the vacuum energy signalling vacuum instability. 
In the present case, there is no vacuum instability and the non-convergent integral over the negative eigenmode is simply
cut-off at a finite point to yield the unknown but finite factor $M$, which is then abosrbed in the definition of $J$.
Note that appropriate factors of $vg$ have been inserted in (\ref {vacuum}) to make $J$ and $d$ dimensionless. 

\section {The Gap Equation}

The interaction
of a propagating quark with a single background sphaleron can be thought of as a mass insertion 
$m(p) /p^2$ \cite {as}. Summing the effect of a dilute gas of sphalerons amounts to modifying the
massless propagator to $i/[A(p)\gamma _{\mu}p^{\mu} - m(p)]$. This gives a self consistent
gap equation whose kernel is the mass insertion due to the single sphaleron:
\be
{m(p) \over p^2} = J\, {\rm {exp}}[ - S_s] d^{N_f}
\left [{m_0 \over (vg)} \right ]^{2N_f}{{\psi_{0}(p)\psi_{0}^{\dagger}}(p) \over m_0} \, .
\label {gapone}
\ee
The r.h.s comes from path integral evaluation of the quark two point function in the zero mode
approximation (averaged over color indices) \cite {as}.
The product of the Fourier transform of the zero modes $\psi_{0}(p)\psi_{0}^{\dagger}(p)$ has been summed over the color
index $a$ and determines the momentum dependence of the mass. Note that once the integral over the collective 
degrees of freedom of the sphaleron are performed the equation in momentum space is a purely
algebraic one. In contrast, the gap equation from the instanton in the $ 3 + 1$ dimensional case
is an integral equation involving an integration over instantons of all sizes.

The form of the zero modes of $i \gamma D$ can be found in ref.~\cite {ringwald}. In the gauge where the sphaleron solution
is given by expression \cite {sphaleron}, the zero mode is:
\be
\psi_{0 a i}(R ) = \epsilon _{a i} N {\rm {\exp }}\left [ -2 \int _0^r {f(R ) \over x dx} \right ] \, 
\label {zeromode}
\ee
where $i$ and $ a$ are spin and color indices respectively and $N$ is a
normalization (unknown, since $f(R )$ is not known in closed form). The Fourier transform of the zero
mode is:
\be 
\psi_{0 } (p) = (vg)^3 \int d^3 x \psi_{0 }(x) {\rm {exp}}(-i p\cdot x) \, .
\label {Fourier}
\ee
Then the expression for $m_0$ becomes
\be
m_0 = {1 \over (vg)^3} \int d p\, m(p) p^2 {{\psi }}_{0 }(p) \psi^{\dagger} _{0}(p) \, .
\label {mzerotwo}
\ee
Writing $m(p) = K_{N_f} (p^2/vg) {\psi_{0}(p)\psi_{0}^{\dagger}}{(p)}$, one gets from equation (\ref {gapone})
\be
K_{N_f} = J {\rm {exp}}[ - S_s] d^{N_f}
\left [ \int dp { (\psi_{0}(p)\psi_{0}^{\dagger}(p))^4 p^4 \over (vg)^5} \right ]  ^{2N_f -1} K_{N_f}^{2N_f -1 } \, .
\label {gaptwo}
\ee
To simplify notation we define 
\be 
C = \left [ \int dp {({\psi_{0}(p)\psi_{0}^{\dagger}}(p))^2 \over (vg)^5} p^4  \right ] \, , 
\label {cone}
\ee
\be 
L = J\, {\rm {exp}}[ - S_s] \, .
\label {lone}
\ee
Then the gap equation has, apart from the trivial solution, the non-trivial solution 
\be
K_{N_f} = {[Ld^{N_f}C]^{1 \over 2(1-N_f)} \over C} \, .
\label {gapthree}
\ee
Equation (\ref {gapthree}) 
is also valid for half-integral values of $N_f$ for which
a parity breaking mass term is expected. Indeed the case $N_f = 1/2$ is interesting because the 
gap equation reduces to a single sphaleron approximation. (Theories with half integral
$N_f$ are ill defined due to the existence of global anomalies \cite {anomaly}. One can however add very heavy
``spectator quarks'' to cancel anomalies). The dynamical mass is found from the relation:
\be
K_{1/2} =  {Ld^{1 \over 2}}
\label {kone}
\ee
$N_f = 2$ is the next interesting case to consider. One finds that no non-trivial solution 
exists for this case unless $F \equiv [Ld^{N_f}C] = 1$. But the quantities $d, C$ and $L$ depend
on the parameters $v, g$ and $\lambda$ and so, one expects, $F$ should depend on the dimensionless
quantities $v/g$ and $\lambda /vg $. 

Using the relation (\ref {kone}) one otains in the arbitrary $N_f$ case:
\be
K_{N_f} =  K_{1/2} \left [ {K_{1/2}^2 C \over L} \right ] ^{ 2N_f -1 \over 2(1-N_f)}
\label {gapfour}
\ee
from where the large $N_f$ limit is obtained to be:
\be 
K_{\infty } = K_{1/2} \left [{ L \over {K_{1/2}^2 C }} \right ] \, .
\label {largen}
\ee

It is necessary to know for what range of these parameters is the semiclassical
formula derived above a good approximation. 
The average sphaleron density must be small for the dilute gas approximation to hold.
In reference \cite {sphaleron}, several {\it ansatze} for the sphaleron were used to compute its
action. The results there indicate that the action of the sphaleron is $ O (v/g)$ with a
relatively weak dependence on ${\lambda \over vg}$. 
The average sphaleron density is found by 
maximizing $<0|0>_n$ with respect to $n$. This gives the most likely value $\bar {n}$ of $n$ to be:
\be 
\bar {n} = {\left [ J {\rm {exp}} [-S_S] d^{N_f} \left [{{m_0} \over (vg)} \right ]^{2N_f} [ (vg)^3 \int d^3 x] \right ]} \, 
\label {barn}
\ee
whence the mean sphaleron density $\rho $ is obtained to be
\be
\rho = {\bar {n} \over \int d^3 x} =  J {\rm {exp}} [-S_S] d^{N_f} \left [{{m_0} \over (vg)} \right ]^{2N_f} (vg)^3 \, .
\label {density}
\ee
The dilute gas approximation holds if the mean sphaleron density multiplied by the
sphaleron size ($ \sim (vg)^{-3}$) is small. Substituting the expression for $m_0$ we get
\be
\rho (vg)^{-3} = \left [ L (Cd)^{N_f} \right ]^{{1 \over 1 - N_f}} << 1 \, .
\label {dilute}
\ee
Computing any term on the r.h.s of equation (\ref {dilute}) requires exact knowledge of the 
sphaleron solution. Lacking that, we can still argue that the condition can be satisfied
for particular ranges of parameters. Consider the large $N_f$ limit. In this limit, 
$\rho \sim (Cd)^{-1}$. Note that $C$ depends only on the coupling ${\lambda \over vg}$ which we hold
fixed. 
Because  $d$ is the (normalized) product of the squares of 
the eigenvalues of the operator $i \gamma D$ in the sphaleron background and the normalization
depends only on $g$, for fixed $g$, $d$ is a function of $v/g$. The natural scale 
of the eigenvalues associated with the sphaleron background is $vg$, hence the eigenvalues
increase if the charge $g$ is kept fixed and $v$ is increased, which is to say that their product
$d$ is an increasing function of $v/g$. But $d$ is normalized so that $d (v=0) = 1$. Therefore 
$d \ge 1$. 
It is not possible, without a more detailed calculation to say if
$\rho $ can be indefinitely large, or to say how large must $v/g$ be for 
the dilute gas approximation to be valid. However it is interesting to note that the dilute
gas approximation may be valid in both the 
regimes $0 < v/g \le 1$ and $1 < v/g$. In the first case, it is known by $1/N_f$ expansions that 
the low energy theory (after
integrating out the massive gauge fields) may break flavor symmetries due to strong four-fermi
interactions. The sphaleron method is more trustworthy in the 
second case, where the four-fermi interaction is weak, and one does not find chiral
symmetry breaking in $1/N_f$ expansions. 
Note that the dynamical mass obtained in (\ref {gapfour}) has the peculiar feature of becoming a 
constant for large $N_f$.

\section {The Sphaleron and Quark Zero Modes}

We have shown that in the $SU(2)$ broken-color theory, sphalerons 
induce parity breaking mass in the $N_f = 1/2$ theory and may break chiral symmetries
for arbitrary $N_f$. Some other questions one may ask at this point are: 
(i) Is the result valid for other color groups? (ii) Is it valid if color is unbroken?
(iii) Are there objects other than the sphaleron which can 
induce chiral symmetry breaking?

The first question can be immediately answered. Suppose we have an $SU(N_c)$ gauge theory ($N_c > 2$) which 
is broken to $SU(N_c-1)$ by a Higgs in the fundamental representation. Then one can take three of the 
$2N_c-1$ broken generators to form an $SU(2)$ in $N_c$ different ways and get $N_c$ distinct sphalerons
that suffice to give mass to all the quarks.
Note that there is an unbroken $SU(N_c-1)$ whose massless gauge bosons
also give a quark condensation effect that vanishes
for $N_f > N_{fc}$. 

Let us now consider the second question. First we summarize what we know about different
limiting cases:\\
(i) When $v/g >> 1$, sphaleron gas is a good approximation and seems to give an order $vg$ mass
to quarks for large $N_f$. \\
(ii) When $v/g \sim 1$, the sphaleron effect may or may not be trustworthy
depending on the validity of the dilute gas approximation. 
However, upon integrating out the massive gauge bosons, one obtains a strongly coupled
theory of four-fermi interactions (interaction strength $G \sim g^2 /(gv)^2$). 
This leads to an order $vg$ mass for quarks when $N_f < N_{fc}$. \\
(iii) When $v/g << 1$, the four-fermi theory is no longer a good
low energy theory because the gauge fields are too light. 
The dilute gas approximation for the sphaleron also breaks down.
Since $g^2 >> vg$,
the gauge fields can be treated as massless. In this case, an order $g^2$
quark mass is dynamically generated by gluon exchange. Like case (ii), this mass
vanishes as $N_f$ approaches $N_{fc}$ from below. 

There is one picture that is consistent with each of the above limiting cases. 
One can think of the quark mass as made of two parts. There is a part that comes 
from light gluon exchange and is order $g^2$. This mass vanishes for large $N_f$. The second
part is an order $vg$ part that comes from sphaleron effects and
approaches a non-zero constant value for large $N_f$. The picture is heuristic since there is
no region of parameters where controlled calculation of all three dynamical effects described above can 
be performed. 

The $N_f = 1/2$ case is interesting because it does not require a gap equation 
or the dilute gas approximation. In this case the quark mass is order $vg$ and
goes to zero as $v \to 0$. Also, for $v=0$ 
no sphaleron exists, indicating that the answer to question (ii) may be ``no'' unless
the sphaleron method can be generalized to work when $v=0$. This brings us to a discussion of question (iii).

Consider a parity invariant theory with two quarks. 
Recall that 
the trace of the quark two point function with the background gauge field $W$ can be written as
\be
\int d^3 x {\rm {Tr}}[S(x, x)]_{W} = \left [\sum _{\lambda } { 1 \over \lambda + im} - {1 \over  \lambda - im} \right ]
= \sum _{\lambda \ge 0} \left [ {2im \over \lambda ^2 + m^2} \right ] \, 
\label {twopoint}
\ee
where $\lambda $ are the eigenvalues of the Dirac operator and $\psi _{\lambda }$ are the corresponding
eigenfuctions. Note that we have taken a parity invariant trace over two quarks with positive and negative 
mass terms. In the chiral limit $ m \to 0$, the sum over $\lambda$ will be zero unless 
there was an infrared divergence. Hence it is
the eigenvalues close to zero which lead to quark condensation \cite {eigen}.

Because a sphaleron has a normalizable
zero mode, a dilute gas of sphalerons has a large number of almost zero modes, with the
eigenvalues vanishing for infinite dilution. 
Gauge field configurations that have special, normalizable 
zero modes like the sphaleron are most suitable for writing a gap equation. Let us denote their
space to be $S_0$. 
What do we know about $S_0$? 
Recall the level crossing picture of the chiral anomaly of 4- dimensions \cite {cdg, level}. Euclidean gauge field configurations
with winding number $n$ contribute to fermion number violating processes by $n$ units. 
Consider the static energy eigenvalue equation:
\be
\gamma_i D_i (1 - \gamma_5) \psi (x_i) = E \psi (x_i) \, \, \, \, (i = 1, 2, 3)
\ee
where the $\gamma $ matrices are four dimensional and there is a background gauge field $A_i (x_\mu),
(\mu = 1, 2, 3, 4)$
of winding number $n$. 
The level crossing picture tells us that as $x_4$ is changed from $-\infty $ to $+\infty$, $n$ 
negative energy levels cross the zero energy level and become postive energy levels (implying a creation of
$n$ quarks). Thus in the instanton background ($ n = 1$) there is a value of $x_4$ ($x_4 = 0$)
for which a normalizable zero eigenmode of the three dimensional Dirac operator exists. 

In the $2+1$ dimensional gauge theory at hand take a one parameter family of field configurations $A_i(x_\mu )$
that begin ($x_4 \to \infty $) and end ($x_4 \to -\infty $) at the vacuum. This is a loop in the space of 
field configurations which is non-contractible (an NCL) if $\int d^4 x F \wedge F = n \ne 0$. The winding number of
this loop is just like the winding number of an $n$ instanton. Therefore there is level crossing as described 
above, and for $n \ne 0$ the loop must intersect the surface $S_0$. 

{\centerline {{\epsfxsize=3.0in\epsfbox{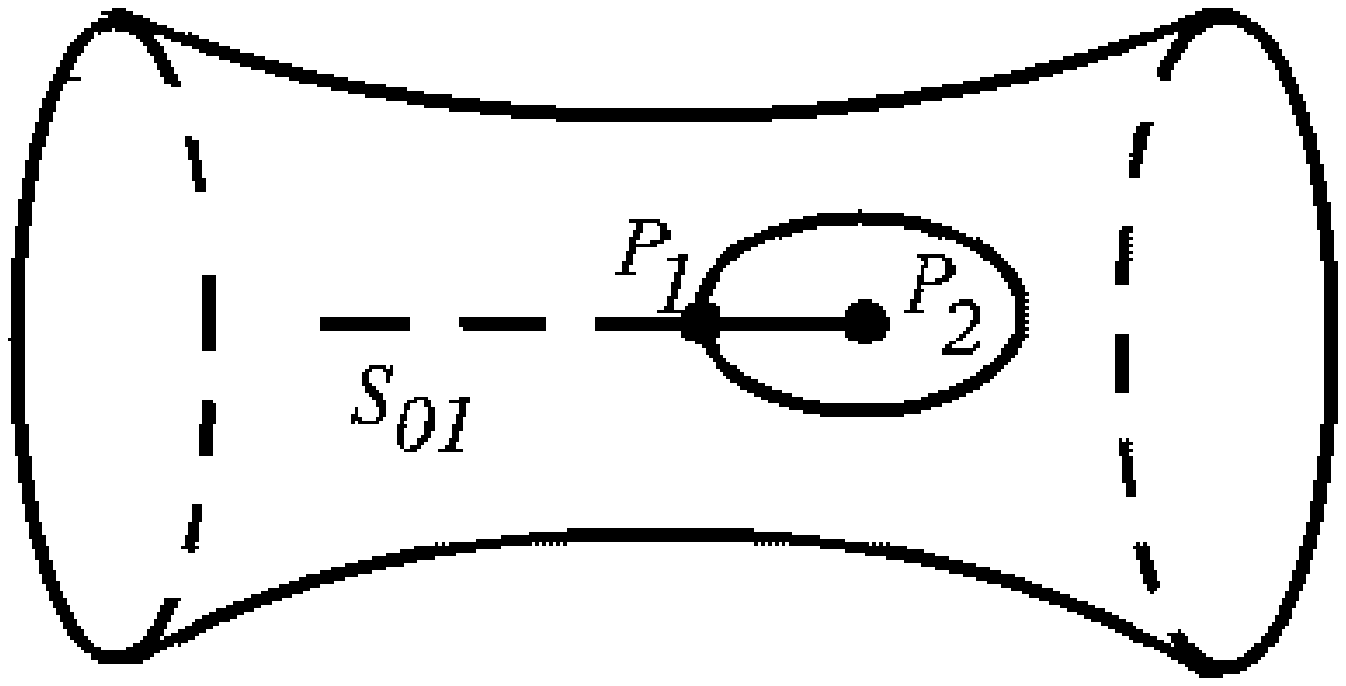}}}}
{\centerline {\makebox[0.8in][l]{\hspace{2ex} Fig. 1.}
\parbox[t]{4.8in}{ {\small The schematic picture shows a torus-like space
that is not simply connected. The co-dimension 1 surface $S_{01}$ is a line
here with a point-like boundary at $P_2$. The circle centered at $P_2$ and passing through $P_1$
intersects $S_{01}$ only once.
}}}}

\vspace {1.0 cm}

Since every NCL must intersect $S_0$, $S_0$ has a part which is a co-dimension 1 surface in the space of 
all field configurations
of finite action. This surface $S_{01}$  must have no boundaries which is proven as follows. 
Suppose $P_2$ is a point on the boundary (Fig. 1). Then there can be a contractible circle
with center at $P_2$ that intersects $S_{01}$ only at a single point $P_1$ which is not
possible because level crossing does not occur around contractible loops. 
Thus $S_{01}$ is an unbounded surface that intersects every NCL. How many such unbounded surfaces are there? Suppose
their number is $s$. Since every
NCL must cross each of these surfaces, one finds $s$ by considering
a particular NCL. The instanton
defines a special NCL with a highly symmetric form for the gauge fields. It is known that
only the $x_4 = 0$ cross section of the instanton has a zero mode (and exactly
one zero mode) \cite {cdg,instanton}. Hence $s = 1$. 
The justification for a saddle point expansion around the sphaleron is that it is likely to 
be the least action point on $S_{01}$ at least for small Higgs mass. Indeed for $M_H < 12 M_W$
the sphaleron is the least action spherically symmetric saddle point of the action \cite {yaffe}.

We can now also answer question (iii).
Firstly, for large Higgs mass the minimum action point on $S_0$ is not the sphaleron
but some other non-trivial field configuration \cite {yaffe, klinkhamer2}.  
Furthermore, if the 
Higgs potential $U$ is minimized at $\phi = 0$ (or if the Higgs does not exist in the theory), 
the sphaleron is no longer
a saddle point of the action. 
For a given configuration of the gauge field $W$, the action is minimized by 
taking $\phi \equiv 0$, and the action can be lowered indefinitely by scaling
$x \to \alpha x, \, W \to 1/\alpha W$ for $\alpha > 1$. Thus the minimum action point on $S_0$ does
not exist. It may still be possible to extract the contribution of $S_0$ by the constrained
instanton approach \cite {affleck}. 
Namely, one can fix a scale for field configuration $W$, minimize the action keeping
the scale fixed, find the contribution of this point to the quark condensate and then integrate over 
the scale. Formally this amounts to finding the contribution of a line in the space $S_0$. This is an
interesting idea that can be the topic of future research in this area. 

The sphaleron-instanton parallel may seem a coincidence, since the topological 
resons for their existence are somewhat different. However, there are objects 
in $3+1$ dimensional gauge theories that are closer parallels of sphalerons. 
In an $SU(2)$ gauge theory these are called the $I^{*}$ instantons \cite {klinkhamer}.
In reference \cite {boomeron}, sphaleron like extrema of the Euclidean action were
dubbed boomerons and a list of theories were they may exist was presented. The 
present results indicate a possible role of boomerons in chiral symmetry breaking. 
Whether $I^*$ plays a similar role in $3+1$ dimensions is yet to be investigated. 

\section {Conclusions}

In conclusion, we have shown that quark mass generation by instanton effects may be
possible in $2+1$ dimensions. The role of the instantons is played by sphalerons.
The unifying idea behind the instanton effects is the existence of a space of
gauge field configurations with a normalizable quark zero mode. If the minimum action
point of this space is well defined, a semiclassical expansion about that point
can lead to a gap equation. Non-trivial solutions of the gap equation signal 
quark condensation and chiral symmetry breaking. The most interesting aspect of
the new dynamical mass is that it vanishes with the color symmetry breaking breaking
parameter $v/g$ but approaches a constant value for large $N_f$. The effect is inherently
non-Abelian and is absent in massive QED.

\centerline {\bf {Acknowledgements}}

We thank Frans Klinkhamer, Sekhar Chivukula and Igor Shovkovy for 
discussions and suggestions. This work was supported by the Department 
of Energy under the grant DE-FG02-84ER40153.

\vfill

\end{document}